\documentclass{article}

\usepackage{arxiv}
\usepackage{amsmath}
\usepackage{graphicx}
\usepackage{subfig}
\usepackage{algorithm}
\usepackage{algorithmic}
\usepackage{listings}
\usepackage{bm}
\usepackage{hyperref}       
\usepackage{url}            
\usepackage{booktabs}       
\usepackage{amsfonts}       
\usepackage{nicefrac}       
\usepackage{caption}

\title{Learning Representations of Hierarchical Slates in Collaborative Filtering}

\author{
  Ehtsham Elahi \\
  Netflix Inc.\\
  100 Winchester Circle\\
  Los Gatos, CA 95032 \\
  \texttt{eelahi@netflix.com} \\
   \And
 Ashok Chandrashekar \\
  Netflix Inc.\\
  100 Winchester Circle\\
  Los Gatos, CA 95032 \\
  \texttt{achandrashekar@netflix.com} \\
}

\begin{document}
\maketitle
\begin{abstract}
We are interested in building collaborative filtering models for recommendation systems where users interact with slates instead of individual items. These slates can be hierarchical in nature. The central idea of our approach is to learn low dimensional embeddings of these slates. We present a novel way to learn these embeddings by making use of the (unknown) statistics of the underlying distribution generating the hierarchical data. Our representation learning algorithm can be viewed as a simple composition rule that can be applied recursively in a bottom-up fashion to represent arbitrarily complex hierarchical structures in terms of the representations of its constituent components. We demonstrate our ideas on two real world recommendation systems datasets including the one used for the RecSys 2019 challenge. For that dataset, we improve upon the performance achieved by the winning team's model by incorporating embeddings as features generated by our approach in their solution.
\end{abstract}



\maketitle

\section{Introduction}
\begin{figure}[!h]
\centering
\includegraphics[width=0.40\columnwidth]{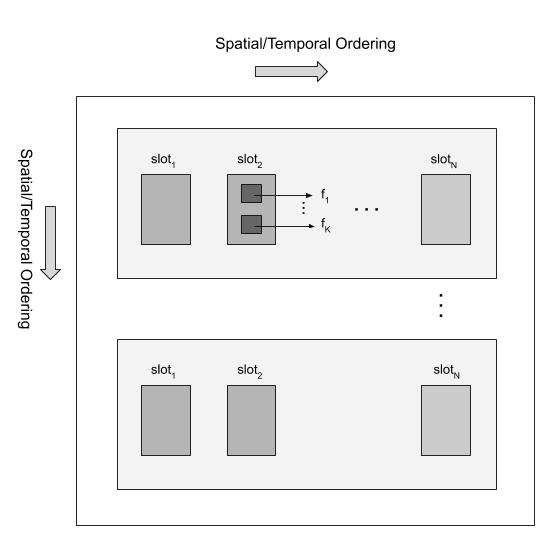}
\includegraphics[width=0.40\columnwidth]{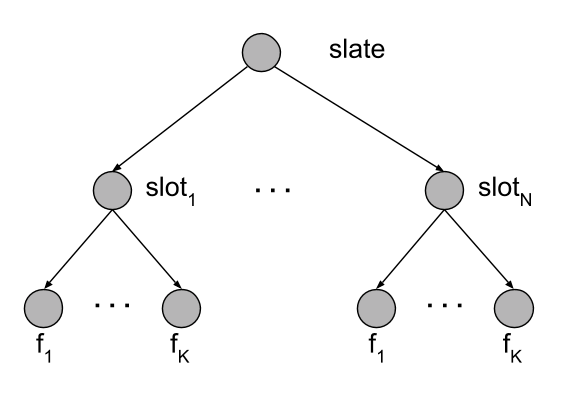}
\caption{A hierarchical slate with multiple sub-slates each with N slots. Each slot has K features $f_1, ..., f_K$. Using a tree is another useful approach to represent such hierarchical slates.}
\label{hierarchical_slate}
\end{figure}

The term slate is widely used in machine learning research to denote an assortment of individual items \cite{NIPS2017_6954}. In this work, we study problems that involve Hierarchical Slates, i.e, when slates themselves are organized in a spatial hierarchy or a temporal sequence or both (See  Figure-\ref{hierarchical_slate} ).
E-Commerce, music and video streaming, travel booking and news websites are examples of applications that can present their users with such hierarchical slates. In video streaming websites for example, customers often see videos arranged in a two dimensional grid to form the entire home page. Such an assortment gives rise to spatial hierarchy. Temporal sequence of items/slates presented to the customer in an interactive recommender system in a session can also be modeled as hierarchical slates. When a customer is presented with such a slate, we may observe feedback from them indicating positive or negative engagement with the entire slate (as opposed to an individual item on the slate). In this work, we are interested in modeling user response to slates and the core to our modeling technique is learning embeddings of the hierarchical slates. While a lot of research has been conducted in learning item embeddings via collaborative filtering techniques \cite{pmf} on user feedback data, not much work has gone into extending the task of representation learning to higher order entities such as slates. These embeddings can benefit learning tasks in sparse high dimensional domains while revealing the underlying structure of the problem space. To summarize,
\begin{itemize}
    \item We consider hierarchical slates which are composite data structures of items arranged spatially (such as a grid) or temporally leading to a hierarchical organization.
    \item We propose a novel method to learn low dimensional representations of hierarchical slates. We call these representations as slate embeddings. 
    \item We propose a novel method to learn low dimensional representations of hierarchical slates presented to a user by recommender systems. We call these representations as slate embeddings. The embedding learning technique is based on summarizing the slates using the first and 2nd order statistics of its constituent elements. 
    \item The embeddings are learned in a supervised fashion via collaborative filtering by optimizing a single aggregate loss function for the dataset.
\end{itemize}

\section{Model}
Let $X = [x_1, …, x_L]$ be a list of $L$ $k$-dimensional random vectors. Let $X$ be a multivariate normal (and so are the marginal distributions for $x_i$, i = 1...L). Let $\mu$ be the mean of Pr(X) and Cov(X) be its covariance matrix. 
$\mu = [\mu_1; … ; \mu_L]$ is $L \times k$-dimensional and Cov($X$) is $(L\times k) \times (L\times k)$ dimensional matrix. The $i,j$ sub-block in Cov(X) = Cov($x_i, x_j)$ and is $k \times k$ dimensional. Given a dataset of $N$ such lists ${X_1, …, X_N}$, $\mu$ and Cov can be estimated using maximum likelihood \cite{bishop} as $\mu = \frac{\sum_i^N X_i}{N}$ and $\text{Cov} (x_i, x_j) = \frac{\sum_{n=1}^N (x_i x_j^T)_n}{N} - \mu_i \mu_j^T$

To connect the above idea to our hierarchical slates representation, let $X$ be a slate of items and $x_i, i=1...L$ are the (unknown) representations/embeddings for the items in the slate. We would like to use a representation of $X$ that makes use of the above two statistics that identify the underlying probability distribution. One possible way would be to simply stack the two statistics 
\begin{eqnarray*}
em(X) = [ \mu_1, …,\mu_L; \text{Cov}(x_1,x_2), …, \text{Cov}(x_{L-1}, x_L)]
\end{eqnarray*}

In practice, it may be too costly to directly use $em(X)$ as that would be very high dimensional $(L\times k) \times (1 + L \times k)$. Therefore we use mean as fast dimensionality reduction on $em(X)$. 
\begin{eqnarray*}
em(X) = [\text{mean}(\mu_1,...,\mu_L); \text{mean}(\text{Cov}(x_i, x_j)\  \forall\  i,j = 1...L) ]
\end{eqnarray*}

Furthermore, we only keep the unique cross-covariance matrices in the 2nd component and drop the covariance terms as we expect to learn the most from co-occurrence of items captured by the cross-covariance sub-blocks. We further only use the diagonal of cross-covariance matrix as another step to reduce dimensionality of cross-covariance term from $k^2$ to $k$. Finally, we can optionally choose to stack the mean and cross-covariance terms $(2 \times k$ dimensional embedding) or add the two (k dimensional embedding for X). Experimentally, we found both to perform similarly therefore we go with the version where we add the two statistics for its smaller dimensionality.  The final representation of the em(X) is 
\begin{eqnarray*}
em(X) =  \text{mean}(\mu_1, …, \mu_L) + \text{mean}(\text{diag(Cov}(x_i, x_j))\ \forall \ i > j)
\end{eqnarray*}

Since both $\mu$ and Cov are unknown, we plug-in the single sample estimates of $\mu$ and Cov($x_i, x_j$). For the cross-covariance term, the single sample estimate would be zero (by subtracting off the means) hence we don’t subtract off the means and only keep the outer-product term. With these plugin estimates, the final representation for X in terms of observed data becomes
\begin{eqnarray*}
em(X) = \text{mean}(x_1, .., x_L) + \text{mean}(\text{diag}(x_i x_j^T) \ \forall \ i > j)
\end{eqnarray*}

For hierarchical slates, the idea is to apply this representation recursively. Let $S = [[x_1, .., x_L]_1, …,[x_1, …, x_L]_M]$ be a list of M slates. Then the embedding for $S$ $em(S)$ would be
\begin{equation}\label{main_eq}
\text{mean}(em(X_1),..., em(x_M)) + \text{mean}(\text{diag}(em(X_l) em(X_m)^T)
\forall\ l > m)
\end{equation}

\begin{figure}[!h]
\centering
\includegraphics[width=0.50\columnwidth]{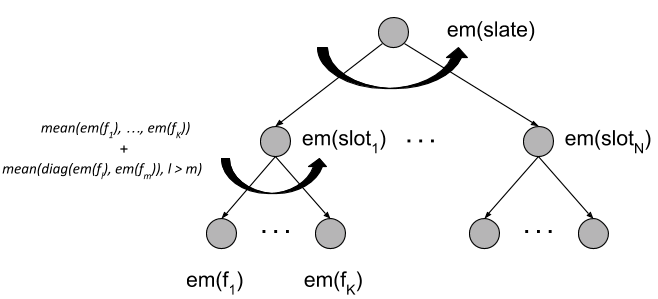}
\caption{A bottom-up recursive application of composition rules to get embeddings of slots and then slate. The curved arrow across the children of a node indicates computing the two statistics to come up with an embedding for the parent node.}
\label{comp_rule}
\end{figure}

And we can extend this idea repeatedly to represent arbitrarily complex data hierarchies. We can simply view the embedding construction as recursively applying the above composition rules on a list of embedding vectors. The computation using the tree representation of the slate would look as shown in figure-\ref{comp_rule}.

The approach is highly scalable and can be implemented in a single pass over the list of embedding vectors to get embedding for parent node. 

As mentioned earlier, our goal is to learn these embeddings in a supervised setting. In collaborative filtering for example, we have users interacting with slates and providing feedback (either explicit ratings or implicit behavioral). We can learn the slate embeddings as one would learn embedding of singleton items in the factorization of user-item matrix \cite{gmf}. For example, the logit function for the user response to a slate can simply be: $\text{logit} = em(X)^T q_u$ where $q_u$ is a $k$-dimensional latent factor for user $u$. We would like to emphasize that our contribution is the construction of slate embeddings and the remaining aspects of the model (logit and loss functions) are dependent on the task at hand and we provide details for these in the experiments section. 

\subsection{How to learn em(f) ?} 
One detail that we have glossed over is how to get the embedding of the features ($em(f_i)$ in figure-\ref{comp_rule} at the leaves of the hierarchical slate tree). Our composition rule starts with the leaf-level embedding vectors and moves bottom-up to learn representations for all the non-leaf nodes of the tree. Clearly, our slate embedding learning approach is independent of how we obtain $em(f)$. The simplest way would be to linearly embed the features in the latent space (like matrix factorization or factorization machine \cite{fact-machine}). That is what we do in the experiments for this work however we can expect to have more powerful models if our slate representation learning algorithms is combined with non-linear techniques to embed features. As part of the learning algorithm, we would learn all embedding vectors for individual items as well as any additional parameters (like $q_u$ in a collaborative filtering task mentioned above). This can be easily done by maximizing the log-likelihood for the data of the task at hand. We can optimize the log-likelihood using gradient based techniques. 

\section{Related Work}
Collaborative filtering is an active area of research with state-of-the-art results regularly published on recommendation systems datasets \cite{slim-harald}, \cite{mult-vae}. However, most of the collaborative filtering research deals with modeling a user’s engagement with a single item (a customer interacting with a movie/song etc.). Gaussian Matrix factorization (MF) and its variants are very popular in collaborative filtering research community \cite{gmf}, \cite{bayesian-mf}, \cite{collective-mf},\cite{10.1145/3298689.3347036}. What our approach shares with matrix factorization is the parameterization of the response/logit function. Like MF, our logit functions use an inner product of slate embedding with user embedding. The main difference of our approach with MF is that we work with slates instead of singleton items. As such, our approach can be viewed as an extension of MF where we apply our simple composition rules to come up with the representation of slates and the rest of the procedure in MF follows. Another technique which is closely related with MF is factorization machine \cite{fact-machine}. Like MF, factorization machine embeds features linearly and then applies pairwise inner-products among embeddings of all features to come up with a response function. Instead of using pairwise inner-products as a vehicle to get the response function, we express the cross-covariance between embedding vectors using pairwise products (outer product for the full cross-covariance or element-wise product for only capturing the diagonal of the cross-covariance matrix). A side effect of this is that we are able to capture up to 4th order interactions among embeddings (i-e when we compute the inner product of embedding vectors, each being represented by pairwise products). In factorization machine expressing a full 4th order interaction among all features is computationally exhaustive.

Any machine learning algorithm is as good as the input features of the data. Our work deals with learning feature representations for slates encountered in collaborative filtering. Similar ideas have been explored under the moniker of learning embeddings or distributed representations in \cite{replearning-1}, \cite{word2vec}. Compared to the popular word2vec algorithm \cite{word2vec}, our approach is supervised therefore the learned representations reflect the supervised task. Moreover, our approach makes use of both numerical and categorical features (word2vec works with categorical features only). 

There is also a large body of work that works with datasets of slates \cite{travel}, \cite{learningfromsets}, \cite{slatebandit}, \cite{NIPS2017_6954}. \cite{learningfromsets} also uses one of the datasets that we have used but their motivation is to improve item level predictions using user's response to slates of recommendations. For slate recommendation problem, \cite{NIPS2017_6954} discusses evaluation techniques and references a few modeling methods as well. 

\section{EXPERIMENTS}

\subsection{MovieLens Slates of Movies dataset}
In the MovieLens slates of movies dataset \cite{learningfromsets}, users are presented with a list of movies and they are asked to provide a single numerical rating for the recommended list. The scale of the rating ranges from 1 to 5 with 5 indicating the most preferred list of recommendations. Summary of the dataset is in table-\ref{movielens-dataset}. An example slate from the dataset is shown in figure-\ref{ml-slate} 

\begin{table}[ht]
\begin{minipage}[t]{0.48\linewidth}
\centering
\caption{Description of MovieLens Slate of Movies dataset}
\begin{tabular}{llll}
\multicolumn{1}{l}{\bf Attribute}  &\multicolumn{1}{l}{\bf Value} \\
\hline \\
Number of items per slate         &5\\
Number of users &854\\
Number of movies &12,549 \\
Number of Training slates         &22,346\\
Number of Validation slates         &3585\\
Number of Test slates         &3585\\
\end{tabular}
\label{movielens-dataset}
\end{minipage}\hfill
\begin{minipage}[t][][b]{0.48\linewidth}
\centering
\frame{\includegraphics[width=65mm]{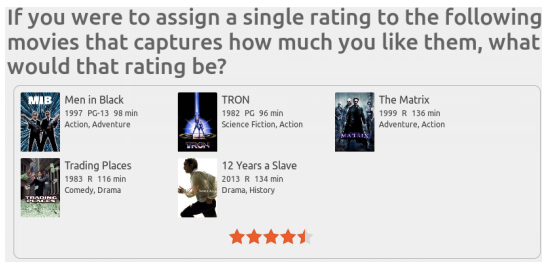}}
\captionof{figure}{An instance of a slate of movies in the MovieLens slate of movies dataset. Customer is asked to provide a response to an assortment of movies instead of a single movie.}
\label{ml-slate}
\end{minipage}
\end{table}

Features for each slot are the categorical identifier for the movie and the integer position. We use the proposed method to come up with slot and slate level embeddings 
\begin{eqnarray*}
em(slot) = \frac{em(movie) + em(position)}{2} + \text{diag}(em(movie) em(position)^T)
\end{eqnarray*}
\begin{equation*}
    em(slate) = \frac{\sum_{i=1}^{5}em(slot_i)}{5} + \frac{\sum_{i>j}diag(em(slot_i) em(slot_j)^T)}{10}
\end{equation*}

In order to model the users' numerical rating, we introduce a normally distributed k-dimensional latent factor vector $q_.$ for each user (exactly like matrix factorization). For user $u$'s numerical rating $r_{un}$ on $n$-th slate $S_n$,
\begin{equation*}
    r_{un} \sim  \mathcal{N}(. |\ q_u^Tem(S_n),\ \sigma^2)
\end{equation*}

and optimizing the log-likelihood $\mathcal{L}$ takes the form of familiar mean square estimation.
\begin{equation*}
 \mathcal{L} = \sum_{u,n} (q_u^T em(S_n) - r_{un})^2
\end{equation*}
We apply an $\ell_2$ regularization on all unknown parameters and optimize the objective using ADAM in TensorFlow. We compute mean square error (MSE) of the held-out set.\\
For experiments, we compare the proposed model with factorization machine, feed forward neural network and gradient boosted decision trees. We use the FastFM \cite{fastfm} implementation of factorization machine technique. We use a 2 hidden layer architecture with sigmoid activation for feed forward neural net. We found sigmoid to perform better than relu for activation functions in this regression task. We use the LightGBM \cite{lightgbm} implementation of GBDT algorithm.
Below we describe the hyper-parameter search for each method,
\begin{itemize}
    \item \textbf{Slate Embedding Model :} Number of latent factors = $[2, 5, 10, 20, 25] \times  \ell_2 = [1e^{-6}, 1e^{-4}, 1e^{-1}, 10, 100]$. Best Model: $(5, 1e^{-4})$. 
    \item \textbf{Factorization Machine:} Number of latent factors = $[2, 5, 10, 20, 25]\  \times\ \ell_2 = [1e^{-6}, 1e^{-4}, 1e^{-1}, 10, 100] \times \ell_2 = [1e^{-6}, 1e^{-4}, 1e^{-1}, 10, 100]$. Best Model: $(20, 10, 100)$.
    \item \textbf{Feed Forward Neural Network: } Layer 1 number of hidden units = $[2,5,10,20,25] \times$  Layer 2 number of hidden units = $[2,5,10,20,25] \times  \ell_2=[1e^{-6},1e^{-4},1e^{-1}, 10, 100]$. Best Model: $(10, 10, 1e^{-4})$.
    \item \textbf{Gradient boosted decision tree :} Number of trees = $[2, 5, 10, 20, 25, 50] \times$  number of leaves = $[2, 5, 10, 20, 25, 50]$. Best Model: $(50, 20)$.
\end{itemize}

Our proposed approach improves upon the baselines (table-\ref{ML-standalone}) however the improvement is within the standard error. Also overall there isn’t a lot of sensitivity in the dataset with respect to model complexity. Figure-\ref{ml-sweep} shows MSE as a function of dimensionality of latent factors for our algorithm. 

\begin{table}[ht]
\begin{minipage}[t]{0.48\linewidth}
\centering
\caption{Comparison of Mean squared error for SEMB, Factorization Machine, Feed Forward Net and GBM on MovieLens Slate of Movies dataset (lower the better). The standard error is around $0.013$}
\begin{tabular}{ll}
\multicolumn{1}{l}{\bf Model}  &\multicolumn{1}{l}{\bf MSE} \\
\hline \\
Slate Embedding Model        &\bf{0.3900} \\
Factorization Machine        &0.4121 \\
Feed Forward Neural Network  &0.4172 \\ 
Gradient Boosted Decision Trees    &0.4165
\end{tabular}
\label{ML-standalone}
\end{minipage}\hfill
\begin{minipage}[t][][b]{0.48\linewidth}
\centering
\includegraphics[width=65mm]{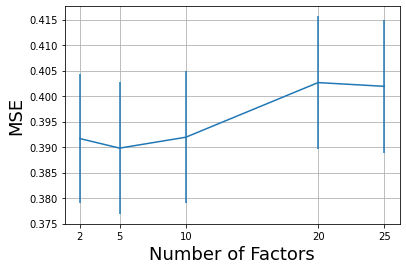}
\captionof{figure}{MSE vs Model complexity for slate embedding model.}
\label{ml-sweep}
\end{minipage}
\end{table}

\subsection{RecSys 2019 Challenge dataset}
In the RecSys 2019 challenge dataset, a customer is recommended a list of accommodations (typically 25) in response to some query (for example customer searching for accommodations at a popular destination). The customer interacts with the presented recommendations over multiple time steps before finally choosing one of the recommendations. The goal of the challenge was to rank the list of recommendations in a successful session. The evaluation metric was mean reciprocal rank \cite{mrr}. Figure-\ref{trivago_data_schema} gives a schematic representation of the dataset and table-\ref{trivago_data_summary} summarizes the dataset.
\begin{table}[ht]
\begin{minipage}[t]{0.48\linewidth}
\centering
\caption{Description of RecSys 2019 Challenge dataset}
\begin{tabular}{llll}
\multicolumn{1}{l}{\bf Attribute}  &\multicolumn{1}{l}{\bf Value} \\
\hline \\
Number of steps per session         &15\\
Number items in slate          &25\\
Number of Train Sessions         &691,463\\
Number of Validation Sessions         &10,000\\
Number of Test Sessions         &125,075\\
\end{tabular}
\label{trivago_data_summary}
\end{minipage}\hfill
\begin{minipage}[t][][b]{0.48\linewidth}
\centering
\includegraphics[width=80mm]{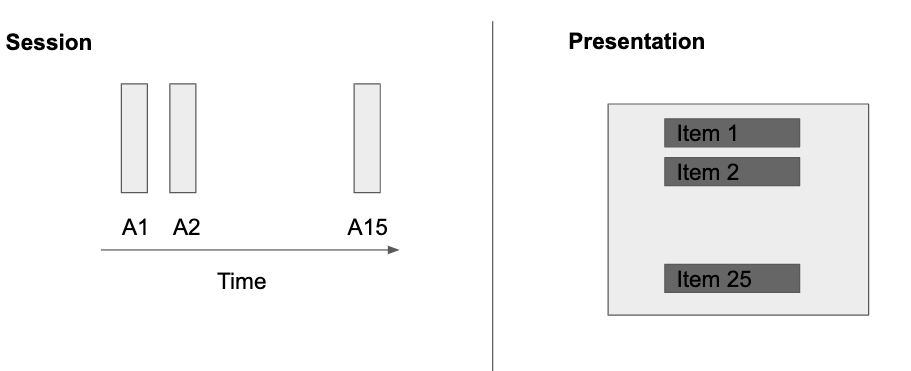}
\captionof{figure}{RecSys 2019 challenge dataset. We get a session view of customers actions over up to 15 time steps (A1-A15). Customers are presented with a slate of recommendations of 25 accommodations}
\label{trivago_data_schema}
\end{minipage}
\end{table}

We follow data processing procedure in \cite{recsys-pub15}. We use 36 features to describe each action and 39 features to represent each item in the slate of recommendations. Both item and action features are summarized in tables \ref{item_features} and \ref{action_features}.

\begin{table}[ht]
\begin{minipage}[t]{0.48\linewidth}
\centering
\caption{Features for items}
\begin{tabular}{ll}
\multicolumn{1}{l}{\bf Feature Indices}  &\multicolumn{1}{l}{\bf Description} \\
\hline \\
0         &Item position\\
1-10  &Item Metadata based features\\
11-13 &User-Item interaction features\\
14-38 &Item price related features\\
\end{tabular}
\label{item_features}
\end{minipage}
\begin{minipage}[t][][b]{0.48\linewidth}
\centering
\caption{Features for actions}
\begin{tabular}{ll}
\multicolumn{1}{l}{\bf Feature Indices}  &\multicolumn{1}{l}{\bf Description} \\
\hline \\
0-9         &10 dimensional 1-hot categorical\\
            &for the action type\\
10-34  &25 dimensional 1-hot categorical\\
            &for the interacted item\\
35          &Time spent in the step
\end{tabular}
\label{action_features}
\end{minipage}
\end{table}

To model the event for a user in session $u$ interacting with an item $i$ in a slate $s$, we construct embedding $em(u)$ of the session by applying equation \ref{main_eq} recursively on features of each action and then on actions across $15$ time steps and embeddings $em(i)$ for all items in the slate $s$ by again applying equation \ref{main_eq} on the features of each item $i$. Furthermore, for each item $i$, we construct an embedding $em(s_i)$ for the entire remaining slate of recommendations by applying embedding construction rule on the set of item embeddings $em(j)\ \forall j \neq i$. 

A simple variant of our model (referred to as SEMB-1 below) would be to consider the interaction of session with an item and ignore the rest of the slate. 
\begin{equation}
    \text{logit}_{u,i} = em(u)^T em(i) \label{eqn2}
\end{equation}
2nd variant (SEMB-2) models the session interaction with the item and rest of the slate as a weighted combination of inner products of the three pairs ($w_1$ and $w_2$ are learnable scalar weights).
\begin{equation}
    \text{logit}_{u,i,s_i} = em(u)^T em(i) + w_1\times em(u)^T em(s_i) + w_2 \times em(i)^T em(s_i) \label{eqn1}
\end{equation}
The multinomial probability of clicking on item $i$ can be obtained by passing the logit through a softmax link function. The log-likelihood $\mathcal{L}$ of the categorical outcomes for the entire dataset is $\sum_{u,i,s_i} \log \Pr(\text{click}_{u,i,s_i})$. We optimize the log-likelihood with an $\ell_2$ penalty on the parameters using ADAM \cite{adam} in TensorFlow. For experiments, we compared the two variants defined in equations \ref{eqn2} and \ref{eqn1} of our proposed approach with factorization machine and feed forward neural network. We extend the vanilla factorization machine model \cite{fact-machine} with a softmax loss function. We did a custom implementation of Multinomial Factorization Machine in TensorFlow. For feed forward net, we used a two hidden layer with relu activation feed forward neural network with a softmax output layer. Details of hyper-parameter selection are 
\begin{itemize}
    \item \textbf{Slate Embedding Model Variant 1 \& 2 (SEMB-1 \& SEMB-2):}  Number of latent factors $=[50, 75, 100, 125]\ \times \ell_2= [1e^{-6}, 1e^{-9}, 1e^{-10}]$. Best model: $(100,1e^{-9})$ for SEMB-1 and $(100, 1e^{-10})$ for SEMB-2.
    \item \textbf{Multinomial Factorization Machine:} Number of latent factors $=[50, 75, 100, 125] \times \ell_2=[1e^{-6}, 1e^{-9}, 1e^{-10}]$. Best model: $(50, 1e^{-10})$.
    \item \textbf{Feed forward Neural Network: } Layer 1 number of hidden units$=[50, 75, 100, 125] \times$ Layer 2 number of hidden units$=[50, 75, 100, 125] \times \ell_2=[1e^{-6}, 1e^{-9}, 1e^{-10}]$. Best Model: $(100, 50,1e^{-6})$.
\end{itemize}
Our approach outperforms the baseline significantly (Table-\ref{trivago_standalone}). It is a rich dataset and responds well to increasing complexity of the model (figure-\ref{semb_sweep}).  
\begin{table}[ht]
\begin{minipage}[t]{0.48\linewidth}
\centering
\caption{Comparison of SEMB-1, SEMB-2, MultiFM and FFWD on the RecSys 2019 Challenge dataset. The standard error is around $0.001$ for MRR results and $0.0008$ for NDCG results}
\begin{tabular}{lll}
\multicolumn{1}{l}{\bf Model}  &\multicolumn{1}{l}{\bf MRR} &\multicolumn{1}{l}{\bf NDCG} \\
\hline \\
SEMB-1         &0.6622 &0.7381\\
SEMB-2         &\bf{0.6640} &\bf{0.7397}\\
Multinomial Factorization Machine          &0.6470 &0.7261\\
Feed Forward Neural Network  &0.6572 &0.7312 
\end{tabular}
\label{trivago_standalone}
\end{minipage}\hfill
\begin{minipage}[t][][b]{0.48\linewidth}
\centering
\includegraphics[width=65mm]{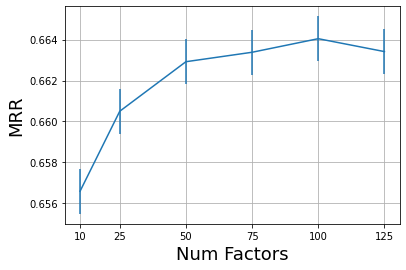}
\captionof{figure}{MRR on the validation dataset with increasing dimensionality of latent factors for SEMB-2}
\label{semb_sweep}
\end{minipage}
\end{table}
\begin{table}[ht]
\begin{minipage}[t]{0.48\linewidth}
\centering
\caption{MRR comparison with features from SEMB-1 and SEMB-2 in the first placed ensemble based solution for RecSys 2019 challenge dataset. The relative improvement over the baseline is $0.058$\% which is similar to the relative improvement in MRR between second and first placed solutions ($0.031$\%)}
\begin{tabular}{ll}
\multicolumn{1}{l}{\bf Model}  &\multicolumn{1}{l}{\bf MRR} \\
\hline \\
Baseline         &0.6829 \\
Baseline + SEMB-1    &0.6832  \\
Baseline + SEMB-2    &\bf{0.6833}
\end{tabular}
\label{ensemble_result}
\end{minipage}\hfill
\begin{minipage}[t][][b]{0.48\linewidth}
\centering
\includegraphics[width=71mm]{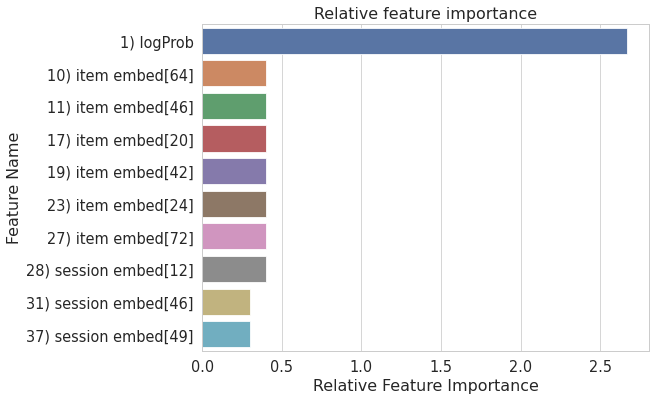}
\captionof{figure}{Rank and Relative Importance of top ten features derived from the SEMB-2 with respect to the top feature in the winning team's solution. For example logProb is the log of the predictive probability at rank 1, item embed[64] is the 64-th element in the item embedding vector at rank 10 in feature importance.}
\label{feature_imp}
\end{minipage}
\end{table}

Furthermore, we add features derived from our model in the LightGBM \cite{lightgbm} based ensemble that won the competition \cite{recsys-pub1}. The winning team's ensemble had over $25,000$ hand-engineered features but inclusion of our model based features improves the performance of the state-of-the-art further (see figure-\ref{ensemble_result}). We also show the feature importance of the new features. The log of the multinomial probability from our model turns out to be the most important feature in the entire ensemble and is more than 2.5x more important than the best feature in the original ensemble (see figure-\ref{feature_imp}).
\subsection{Discussion of results}
We believe that our hierarchical modeling of the positional and temporal dimensions is the reason our slate embedding models improve upon factorization machine and feed forward nets which have a flat structure. Moreover, a side effect of our cross-covariance modeling approach is that we are able to capture upto 4-th order interactions (when an inner product is taken between two embeddings vectors both containing pair-wise interaction terms). In comparison, expressing 4-th order interactions in factorization machine is computationally expensive.

\section {Embedding Visualization and Qualitative analysis}
\begin{figure}[!h]
\centering
\includegraphics[width=0.35\columnwidth]{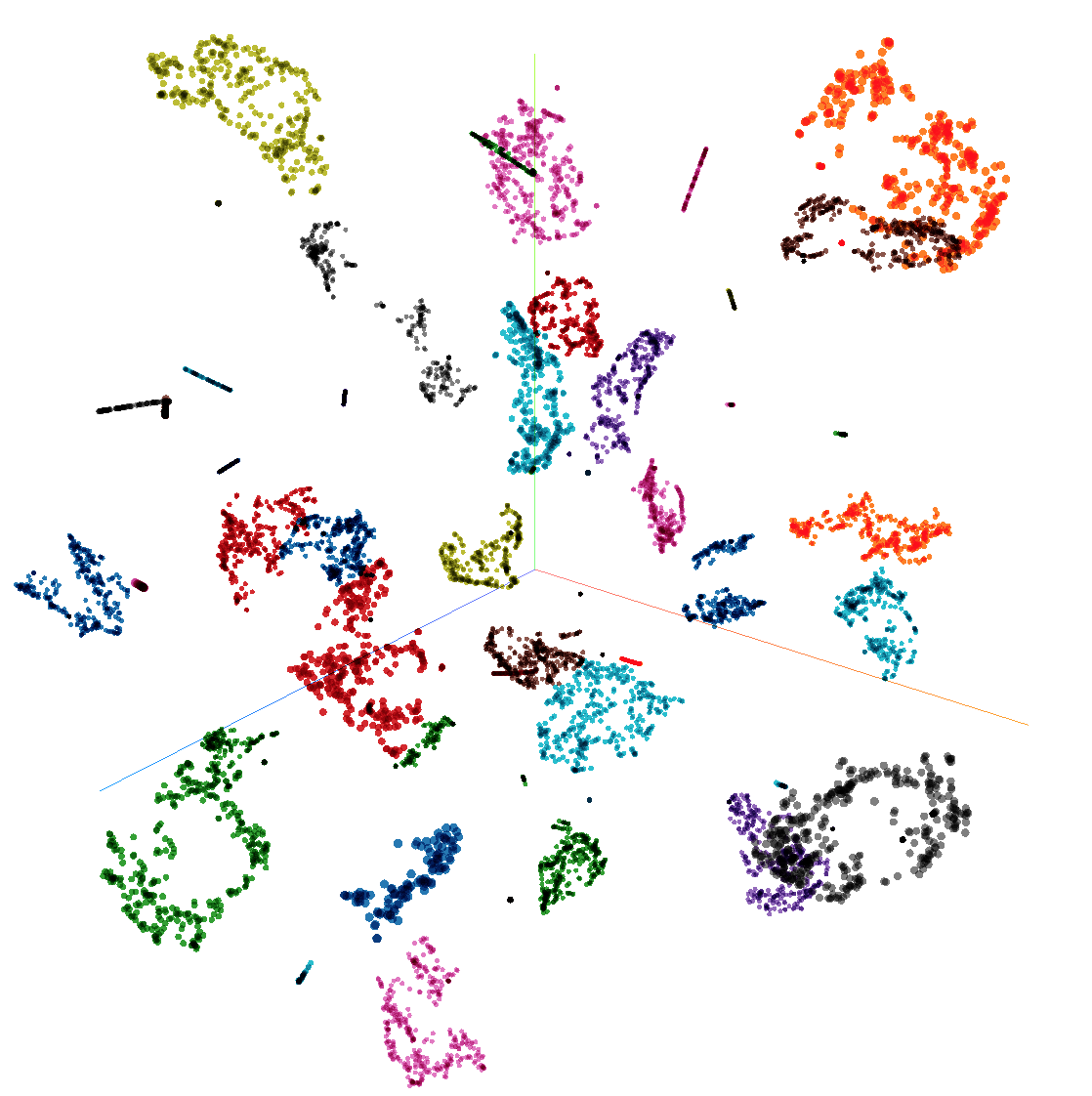}
\caption{Slate Embeddings projected in 3D using t-SNE. Points are colored by the positions of hotels in the slates.}
\label{embed_viz}
\end{figure}

One of the motivations for our work was to be able to visualize hierarchical slates using the embeddings that our model learns. It is important to point out that these embeddings are being learned by a model which is trying to model user's response to these slates therefore any visual structure we see in the data reflects the supervised task that we are solving. In figure-\ref{embed_viz} we show example visualizations from our model. The visualizations was generated by 3D t-SNE projections of 75 dimensional embedding vectors of items (hotel accommodations) in the RecSys 2019 challenge dataset. Visually, the embeddings seem to form many clusters. When we color each point (hotels/accommodation) with its (spatial) position in the slate of recommendations, we find that these clusters closely conform to the 25 positions in the slate. It is not surprising to see the clusters conforming to the positions of items in the recommendation slate. It is well known that recommendation datasets have a high level of presentation bias and this visualization reflects that. 

\begin{figure}[!h]
\centering
\includegraphics[width=0.30\columnwidth]{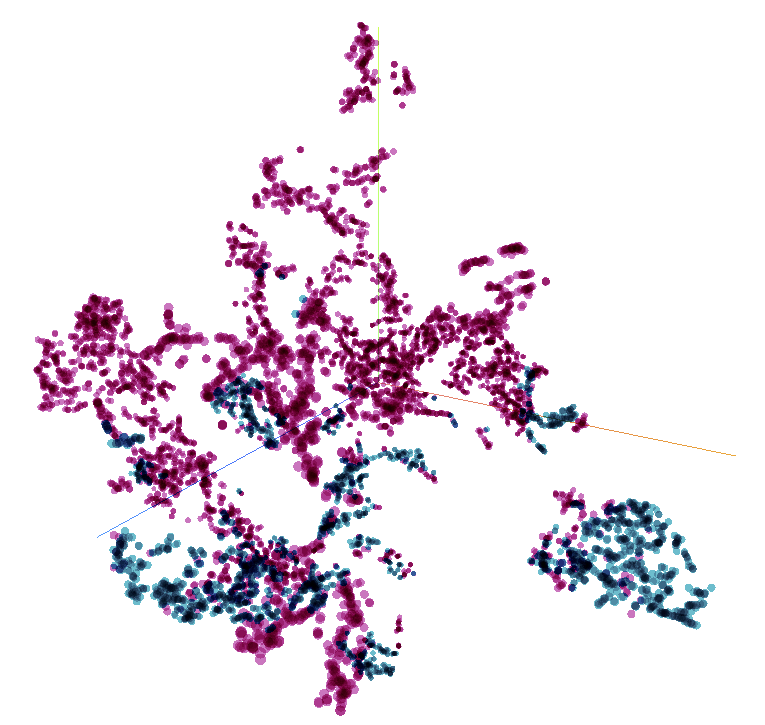}
\includegraphics[width=0.30\columnwidth]{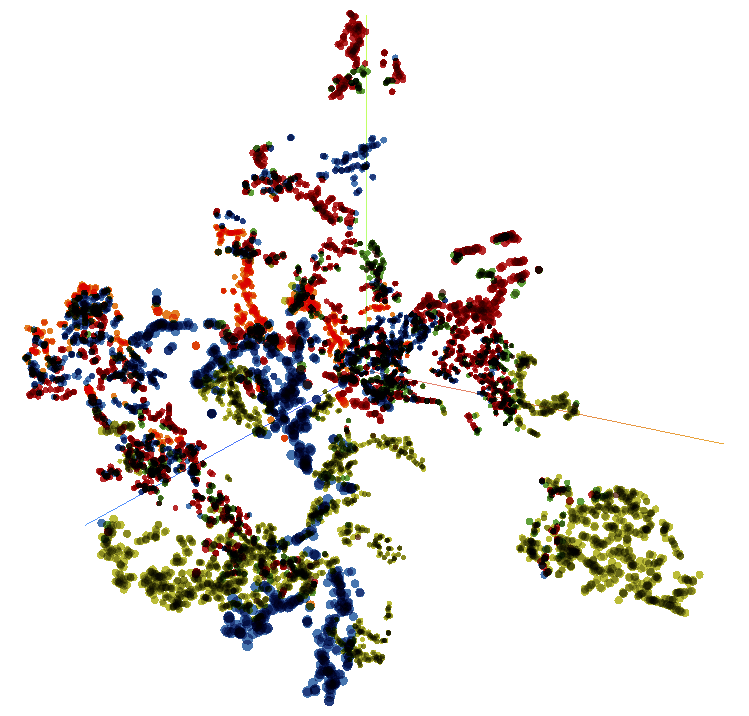}
\caption{These two plots show embeddings for properties presented at the top position in the slate colored by binary indicator is-hotel (left) and star rating of the property (right).}
\label{embed_viz_p0}
\end{figure}

To gain more insight, we focus on only the points presented in the top position in the recommendation slate (figure-\ref{embed_viz_p0}). We color each point with the binary label "is-hotel" (not all accommodations in the dataset are hotels) (left plot) and the star rating of the property (right plot). We again find that even within items presented in the top position of the slate, users' behavior is different as shown by different colored (overlapping) clusters forming. One can spend endless cycles analyzing data using these embeddings to reveal interesting insights.

\section{Conclusion}

In this task, we presented a novel solution to the task of learning embeddings of hierarchical slates. Our approach relied on using the statistics of the underlying distribution generating the hierarchical data. Using well developed principles of collaborative filtering, we use customer feedback on slates to constrain the learning of the slate representations. In the Trivago dataset task, we showed that we can learn these embeddings for hierarchical data structures that have as many as 3 modes of organization (spatial, temporal and feature hierarchies). All it took to learn the embedding of this complicated slates was a recursive application of our technique across all dimensions. We showed that when our approach to representation learning from hierarchical slates achieves competitive performance compared to popular baselines while being simpler. Moreover, by using the ourputs of our framework to augment the hand crafted high dimensional feature representation, we were able to improve on the state-of-the-art solution to RecSys 2019 dataset. The features produced by our model were more than 2 times more influential in the overall ensemble which already had over 25,000 features. While we relied on linear embedding in the leaf level features in the latent space (like matrix factorization), it would be interesting to see how these representations work when we plug them in more powerful deep learning style models to learn non-linear embedding of features. We leave this as future work.

\section*{Acknowledgement}
We would like to thank our colleagues Mehmet Yilmaz, Pannaga Shivaswamy, Henry Wang, Hua Jiang, Jiangwei Pan, Justin Basilico, Chris Steger and Vijay Bharadwaj for valuable discussion and feedback.

\bibliographystyle{unsrt}
\bibliography{reference} 
\end{document}